\documentclass[aps,prl,twocolumn,reprint,superscriptaddress,showpacs,showkeys]{revtex4-1}
\usepackage{graphicx}
\usepackage{amsmath}
\usepackage{xcolor}
\usepackage{siunitx}

\begin{document}

%\preprint{}

\title{Disruption of the $sp^2$ bonding by the compression of the $\pi$-electronic orbitals of graphene at various stacking orders}


\author{Y. W. Sun}
\email{yiwei.sun@qmul.ac.uk}
\affiliation{School of Engineering and Materials Science, Queen Mary University of London, London E1 4NS, United Kingdom}
\author{D. Holec}
\email{david.holec@unileoben.ac.at}
\affiliation{Department of Physical Metallurgy and Materials Testing, Montanuniversit\"{a}t Leoben, Leoben 8700, Austria}
\author{D. N{\"o}ger}
\affiliation{Department of Physical Metallurgy and Materials Testing, Montanuniversit\"{a}t Leoben, Leoben 8700, Austria}
\author{D. J. Dunstan}
\affiliation{School of Physics and Astronomy, Queen Mary University of London, London E1 4NS, United Kingdom}
\author{C. J. Humphreys}
\email{c.humphreys@qmul.ac.uk}
\affiliation{School of Engineering and Materials Science, Queen Mary University of London, London E1 4NS, United Kingdom}
%\thanks{}


\date{\today}

\begin{abstract}
We investigate the behaviour of the $\pi$-electrons under compression and the effect of the stacking order of graphene layers. First we find that electrons can hardly be squeezed through the $sp^2$ network, regardless of the stacking order. The largely deformed electronic orbitals (mainly those of $\pi$-electrons) under compression along the $\textit{c}$-axis increase interlayer interaction between graphene layers as expected, but surprisingly in a similar way for the A-A and Bernal stacking. On the other hand, the large out-of-plane compression shifts the in-plane phonon frequencies of A-A stacked graphene layers significantly and very differently from Bernal stacked layers. We attribute these results to the $sp^2$-electrons filling the low-density central area in a carbon hexagon under compression for the A-A stacking, hence resulting in a non-monotonic change of the $sp^2$-bonding. The results strongly suggest not to ignore 3D features of a 2D material.
\end{abstract}

%\pacs{62.50.-p, 63.20.-e, 63.20.dk, 63.22.Np}
%\keywords{}

\maketitle

Grahene has many extraordinary properties, such as its large in-plane stiffness\cite{Lee08}, mainly due to its featured $sp^2$ network. Multi-layer graphene of each number of layers and stacking order has unique properties, determined by its interlayer interaction, to which mainly the overlap of the $\pi$-electronic orbitals contributes. It is of fundamental importance and interesting to understand and quantify how $sp^2$- and $\pi$-electronic orbitals affect each other. In this work, we study the change of the in-plane properties under uniaxial compression along the \textit{c}-axis and the effect of stacking order, which we expect to have great effects on the behaviour of the $\pi$-electrons.

Researchers usually apply two-dimensional analysis to graphene, a 2D material. In particular, the frequencies of the in-plane phonons of graphene layers were related to only in-plane strain,\cite{Thomsen02,Proctor09,Huang09,Mohiuddin09} despite the out-of-plane strain being about 30 times larger than the in-plane under hydrostatic compression due to the large anisotropy.\cite{Bosak07} In previous work, we quantified the contribution of the out-of-plane strain to the in-plane phonon frequency and found that it could not be neglected. We attributed this contribution to the compression of the $\pi$-electrons into the $sp^2$ network to alter the in-plane bond.\cite{Sun15} To further understand this behaviour, we investigate the effect of stacking order. 

Stacking order has great impacts on the properties of graphene layers. We take the A-A stacking as an extreme example to compare with the normal Bernal stacking. A-A stacked graphene layers are expected to have larger optimised interlayer separation and higher energy than Bernal stacking.\cite{Aoki07} They has some unique electronic/magneto-electronic properties\cite{Lu07}, such as good tunnelling conductance\cite{Popov13} and Fano anti-resonance in the conductance.\cite{Gonzalez10} They also have high optical conductivity in THz range.\cite{Lin12} While most study is theoretical, A-A stacked graphene layers have also been experimentally observed. Lauffer \textit{et al.} observed an area of the A-A stacking in bi-layer graphene by scanning tunnelling microscopy (STM) and spectroscopy (STS)\cite{Lauffer08} and Liu \textit{et al.} found A-A stacked bi-layer graphene close to the folding edge and concluded that the A-A stacking minimised the local strain during the heat treatment.\cite{Liu09} Under compression, for the A-A stacked graphene layers, one would expect them to be easiest to form strong interlayer covalent bond among all the stacking orders. De Andres \textit{et al.} reported an interlayer covalent bond of 0.156 nm after compressing the A-A stacked bi-layer graphene in a theoretical work.\cite{Andres08} We notice that this is a very large compression, beyond the stress range up to 10 GPa (corresponding to interlayer spacing of about 0.23 nm) in this work, and therefore the behaviour of the $\pi$-electrons discussed here does not involve the $sp^2$ to $sp^3$ transition.

\section*{Methods}
We employed density functional theory (DFT)\cite{DFT1,DFT2} as implemented in the Vienna Ab initio Simulation Package (VASP)\cite{VASP} to study the bi-layer graphene and graphite of the A-A and Bernal stacking at 0 K. We treated the exchange-correlation effects by the generalised gradient approximation (GGA) as parameterized by Perdew, Burke and Ernzerhof\cite{GGA} and used the projector augmented-wave method pseudopotentials\cite{PP} for carbon. We used the plane-wave cut-off energy of 900 eV and sampled the reciprocal unit cell with an 18x18x9 k-mesh to achieve the optimised accuracy of the results. We included the effects of Van der Waals (vdW) interaction using the Grimme method\cite{vdW} as implemented in the VASP code. We calculated the vibrational frequencies at the Brillouin zone centre, the $\Gamma$ point, using the 2x2x2 supercell employing the finite displacement method as implemented in the Phonopy code.\cite{phonon}

\section*{Results}
\begin{figure}
	\includegraphics[width=1\columnwidth]{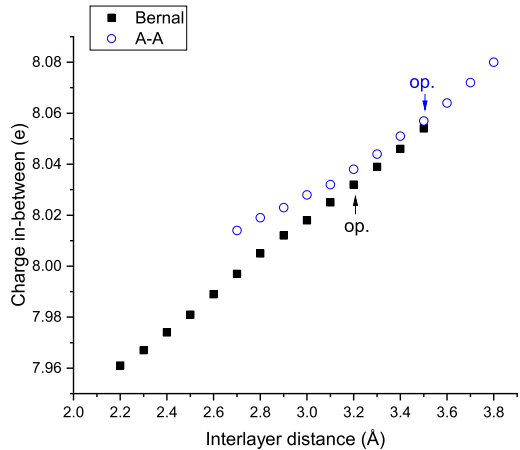}
	\caption{The integrated charge between the two graphene layers is plotted versus the interlayer distance of the bi-layer graphene of the A-A (blue circles) and Bernal stacking (black squares). The optimised interlayer distance for each stacking is labelled.}
	\label{bie}
\end{figure}

We first investigated if the $\pi$-electrons can be squeezed through the $sp^2$ network. We modelled the bi-layer graphene of the A-A and Bernal stacking, varied the interlayer spacing and integrated the charge  between the two graphene layers of the electrons in the outmost occupied shell. For both stacking order, we had 4 carbon atoms in a unit cell and 16 e the sum of the charge from between and outside the two layers. We plot the integrated charge between the layers versus interlayer distance in Fig. \ref{bie}. Compared to Bernal stacking, the optimised interlayer spacing of A-A stacking is larger and it is harder for the electrons to be squeezed through the $sp^2$ network as expected. Nevertheless, for both stacking orders the amount of electrons squeezed through is extremely small. Only 0.53\% of the charge in between the graphene layers is squeezed out under compression of 23\% reduction in volume in the A-A stacking and under a similar compression of 22\% in the Bernal stacking, only 0.63\% of the charge is squeezed out. The consequential large increase of the charge density between the graphene layers under compression indicates a large deformation of the electronic orbitals (mainly of the $\pi$-electrons one would expect). This validates the interpretation in our previous work that the compression of the $\pi$-electronic orbitals is responsible for the significant contribution of the out-of-plane strain to the frequency shifts of the in-plane phonons. Also the smooth change of the charge between the layers suggests that there is no $sp^2$ to $sp^3$ transition.

\begin{figure}
	\includegraphics[width=\columnwidth]{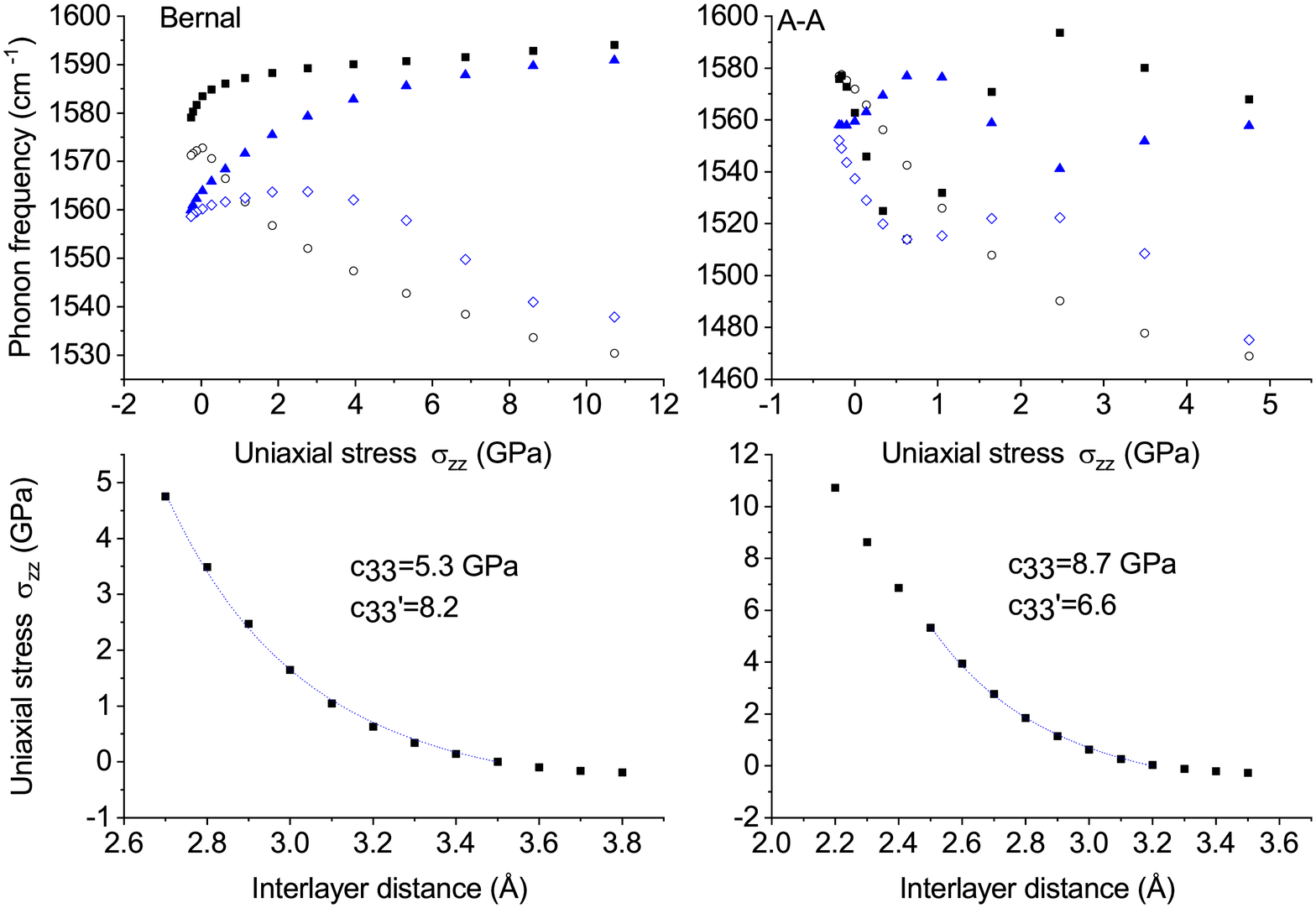}
	\caption{The frequencies of the 4 in-plane phonons of A-A and Bernal stacked bi-layer graphene are plotted with uniaxial stress along the \textit{c}-axis. The solid points are for the two-plane-in-phase modes and the blue ones are for the modes along the C-C bonds. The uniaxial stress was calculated at each interlayer distance as plotted in the lower two sub-figures and the data are fitted by Eq. \ref{m}. The two sub-figures on the left are for the Bernal stacking and the two on the right are for the A-A as labelled.}
	\label{biph}
\end{figure}

We then quantified the effect of the largely deformed electronic orbitals on in-plane and out-of-plane stiffness and the anharmonicity of the A-A and Bernal stacked bi-layer graphene. We applied uniaxial stress along the \textit{c}-axis (the in-plane stress is 0) to the bi-layer graphene and calculated the frequencies of the 4 in-plane phonon modes --- 2 carbon atoms vibrate in-line antiphase along \textit{x} or \textit{y} direction in the hexagonal plane of graphene, and the vibrations in the two layers vibrate in- or out-of-phase. The input in the calculations was the interlayer distance, at which the uniaxial stress was calculated. The results are plotted in Fig. \ref{biph}.    

We fit the data of the uniaxial stress to the interlayer distance by the one-dimensional (along the \textit{c}-axis) analog of the Murnaghan equation\cite{Hanfland89} up to 5 GPa
\begin{equation}
a_{33}/a_{33_0}=[(c_{33}^\prime/c_{33})P+1]^{-1/c_{33}^\prime},
\label{m}
\end{equation}
where $a_{33_0}$ is the unstrained interlayer distance, $c_{33}$ is the elastic constant and $c_{33}^{\prime}$ is the shift rate of $c_{33}$ with pressure. We obtained $c_{33}$=5.3 GPa and $c_{33}^{\prime}$=8.2 for the Bernal stacking, and $c_{33}$=8.7 GPa and $c_{33}^{\prime}$=6.6 for the A-A stacking, compared to the experimental value $c_{33}$=38.7$\pm$0.7 GPa\cite{Bosak07} and $c_{33}^{\prime}$=11.8$\pm$0.9\cite{Hanfland89} of Bernal stacked graphite. The small fitted values of $c_{33}$ and the similar values of $c_{33}^{\prime}$ to graphite, suggest that the bi-layer graphene of both Bernal and A-A stacking are very graphite-like and similar to each other regarding the out-of-plane compressibility. The stacking order does  make a slight difference, and not as expected the A-A stacked bi-layer graphene (with the $\pi$-electronic orbitals from the neighbouring layers largely overlap) becomes softer out-of-plane than the Bernal stacking with increased uniaxial compression. Again the smooth fit by the Murnaghan equation suggests that there is no $sp^2$ to $sp^3$ transition over the presented stress range. 

The shift of the in-plane phonon frequencies of the Bernal stacked bi-layer graphene is understandable that the increasing interlayer interaction under uniaxial compression increases the frequencies of the two in-phase vibrations and generally lower those of the out-of-phase ones. We would like to point out that the shift of the in-plane phonon frequencies under uniaxial compression is comparable to that of graphite under hydrostatic pressure (4.7 cm$^{-1}$GPa$^{-1}$\cite{Hanfland89}), and therefore to consider the effect of the $\pi$-electrons on in-plane properties is desirable. For the A-A stacking, the frequency shifts of 3 out of 4 phonon modes change the sign two times over a small pressure range to 5 GPa, while no $sp^2$ to $sp^3$ transition occurs. This requires further investigation and we study graphite with symmetry on both sides of a graphene layer and have published experimental data to compare with.

\begin{figure}
	\includegraphics[width=0.9\columnwidth]{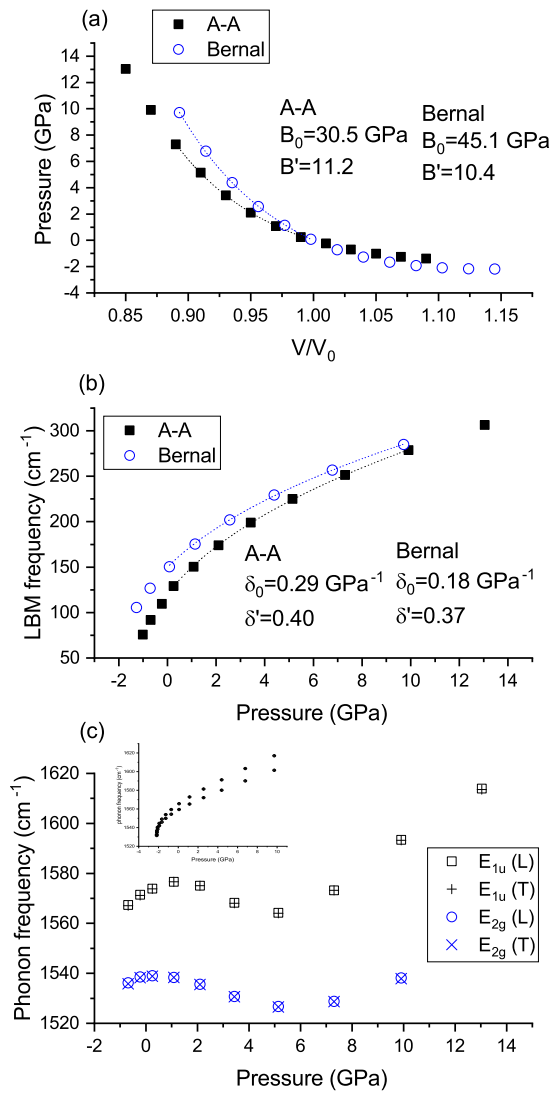}
	\caption{Hydrostatic pressure was applied to the A-A and Bernal stacked graphite. (a) The calculated pressure is plotted versus the ratio of the unit cell volume to the unstrained (the A-A stacking in black solid squares and the Bernal stacking in blue open circles). The fitting by the Murnaghan equation is presented in dashed lines, black for the A-A stacking and blue for the Bernal stacking. (b) The frequencies of the LBM of the A-A and Bernal stacked graphite are plotted versus hydrostatic pressure (the A-A stacking in black solid squares and the Bernal stacking in blue open circles). The data are fitted by Eq. \ref{wp} in dashed lines, black for the A-A stacking and blue for the Bernal stacking. (c) The frequencies of the 4 in-plane phonons modes (as labelled) of the A-A stacked graphite are plotted versus pressure. We use the same notation to label the phonon modes as those 4 in Bernal stacked graphite to be consistent, despite the symmetry is different. L is for the longitudinal modes and T is for the transverse modes. The previously published data of the Bernal stacked graphite is presented for comparison in the inset.\cite{Sun15}}
	\label{nfig1}
\end{figure}

We first modelled graphite of the A-A and Bernal stacking under hydrostatic pressure. We applied pressure by setting a smaller unit cell volume, optimising the geometry and calculating the corresponding pressure. We plot the calculated hydrostatic pressure with the unit cell volume for the A-A and Bernal stacked graphite in Fig. \ref{nfig1} (a). We fit the data by the Murnaghan equation\cite{Murnaghan44} and obtained the unstrained bulk modulus $B_0$=30.5 GPa and its shift rate with pressure $B^\prime$=11.2 of the A-A stacking and $B$=45.1 GPa and  $B^\prime$=10.4 of the Bernal stacking, close to the published experimental values of the Bernal stacked graphite of $B$=33.8$\pm$3.0 GPa and $B^\prime$=8.9$\pm$1.0.\cite{Hanfland89}

The small bulk modulus of graphite is mainly attributed to its weak interlayer interaction, of which the frequency of the layer breathing mode (LBM) is a good indicator. We plot the LBM frequency versus pressure in Fig. \ref{nfig1} (b) and empirically fit the data by\cite{Hanfland89}
\begin{equation}
\omega(P)/\omega_0=[(\delta_0/\delta^\prime)P+1]^{\delta^\prime}.
\label{wp}
\end{equation}
where $\delta_0$ is the logarithmic pressure derivative ($d$ln$\omega$/$dP$)$_{P=0}$, and $\delta^{\prime}$ is the pressure derivative of $d$ln$\omega$/$dP$. We obtained the values of the fitting parameters of $\delta_0$=0.29 GPa$^{-1}$ and $\delta^{\prime}$=0.40 for the A-A stacked graphite and $\delta_0$=0.18 GPa$^{-1}$ and $\delta^{\prime}$=0.37 for the Bernal stacking, compared with the experimental values of $\delta_0$=0.15 GPa$^{-1}$ ($\delta^{\prime}$ not available).\cite{Alzyab88}

For the bulk modulus and the LBM frequency of the Bernal stacked graphite under hydrostatic pressure, the theoretical results are very close to those from experiments, validating the calculations in this work. On contrast to the results from the previous calculations on the bi-layer graphene, despite being initially softer, the A-A stacked graphite stiffens faster than the Bernal stacking with increased pressure as expected. And reasonably the interlayer interaction of the A-A stacked graphite increases faster with pressure, as indicated by the shift of the LBM frequency. However, the different stacking orders only makes a marginal difference in the out-of-plane stiffness and the interlayer interaction, that is mainly determined by the overlap of the $\pi$-electronic orbitals, where the impact of the stacking order ought to be large. Again, no $sp^2$ to $sp^3$ transition occurs in the plotted pressure range.

We then investigated how the stacking orders affect the in-plane properties of graphite. We calculated the frequencies of the in-plane phonons of the A-A stacked graphite and compared them to the published results of the Bernal stacking in Fig. \ref{nfig1} (c). In both stacking orders, the frequencies of the vibrations along the \textit{x}- and the \textit{y}-direction degenerate as expected. The frequencies of both the in-phase and out-of-phase vibrations in the A-A stacked graphite shift non-monotonically with pressure, unlike in the Bernal stacking, over the pressure range where no $sp^2$ to $sp^3$ transition occurs. The compression of the $\pi$-electronic orbitals not only modifies the shift rates of the in-plane phonons with pressure in the Bernal stacked graphite, but also changes the sign of the shifts in the A-A stacking. This is surprising.

\begin{figure}
	\includegraphics[width=0.9\columnwidth]{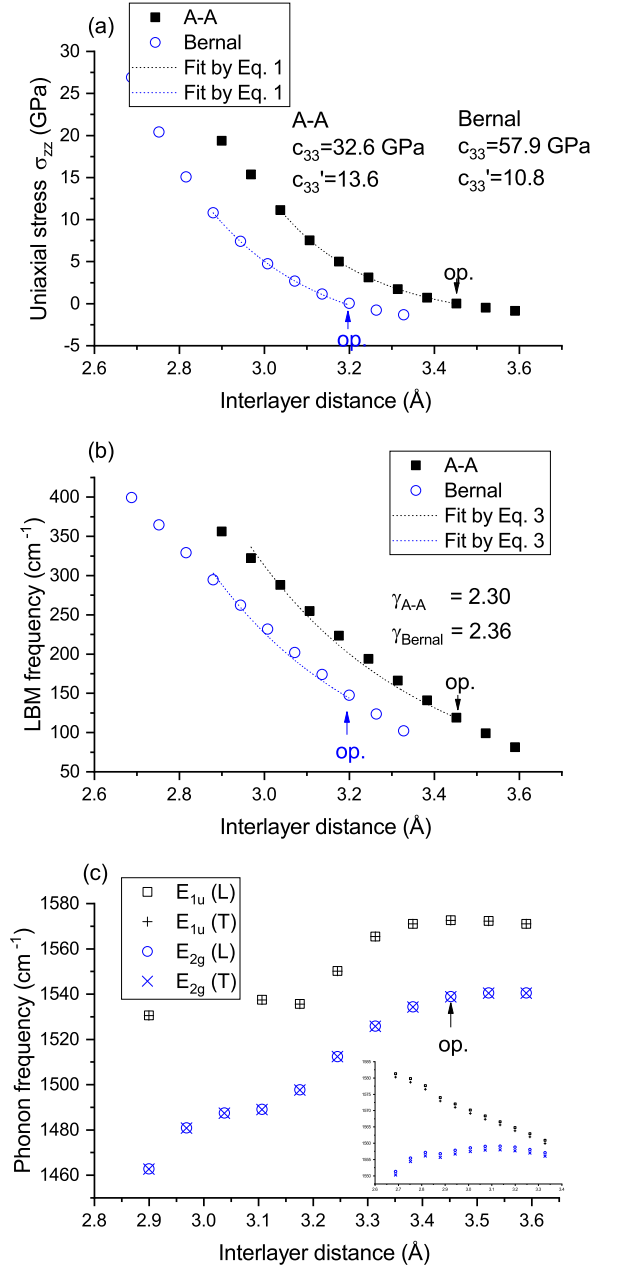}
	\caption{Uniaxial stress along the \textit{c}-axis was applied to the A-A and Bernal stacked graphite. (a) The uniaxial stress was calculated at each interlayer distance of the A-A stacked (black solid squares) and Bernal stacked (open blue circles) graphite. The optimised interlayer distance of each stacking order is labelled. The fit by Eq. \ref{m} is presented in dashed lines, black for the A-A stacking and blue for the Bernal stacking. (b) The frequency of the LBM of the A-A stacked graphite (blue open circles) is plotted versus the interlayer distance. The data are fitted by Eq. 3 in dashed lines. The published data and the fit of the Bernal stacked graphite is presented for comparison.\cite{Sun18} (c) The frequencies of the 4 in-plane phonons modes (as labelled) of the A-A stacked graphite are plotted versus the interlayer distance under uniaxial stress. We use the same notation to label the phonon modes as those 4 in Bernal stacked graphite to be consistent, despite the symmetry is different. L is for the longitudinal modes and T is for the transverse modes. The previously published data of the Bernal stacked graphite is presented for comparison in the inset.\cite{Sun15}}
	\label{nfig2}
\end{figure}

We excluded the effect of in-plane strain by applying uniaxial stress along the \textit{c}-axis to graphite of each stacking order. The interlayer distance was the input in the calculations. We calculated the corresponding uniaxial stress at each interlayer distance and plot the data in Fig. \ref{nfig2} (a). We fit the data by Eq. \ref{m} up to 10 GPa and obtained the $c_{33}$=32.6 GPa and $c_{33}^{\prime}$=13.6 of the A-A stacked graphite and $c_{33}$=57.9 GPa and $c_{33}^{\prime}$=10.8 of the Bernal stacking. The result is consistent with the bulk moduli shown in Fig. \ref{nfig1} (a), that the A-A stacked graphite is initially softer, but stiffens faster than the Bernal stacking. The calculated values of both stacking orders are in general close to the experimental values of Bernal stacked graphite.\cite{Hanfland89} 

We again calculated the LBM frequencies as a measure of interlayer interaction, but here plot them versus the interlayer distance in Fig. \ref{nfig2} (b). We fit the data by\cite{Sun18}
\begin{equation}
\omega(P)/\omega_0=[r(P)/r_0]^{-3\gamma},
\end{equation}
where $\omega$ is the frequency of the LBM and \textit{r} is the interlayer distance. We obtained $\gamma$=2.30 of the A-A stacked graphite, compared with the published $\gamma$=2.36 of the Bernal stacked graphite.\cite{Sun18} The result indicates that the interlayer interaction of the A-A stacked graphite increases nearly as the same rate as that of the Bernal-stacked graphite under uniaxial compression.

Under uniaxial compression along the \textit{c}-axis, the stacking order of graphite has small impact on the out-of-plane stiffness and the interlayer interaction. We then investigated the in-plane properties. We calculated the frequencies of the 4 in-plane phonons of the A-A stacked graphite at each interlayer distance and plot the data in Fig. \ref{nfig2} (c). The published data of the Bernal stacked graphite is presented as the inset for comparison. We notice that the out-of-plane compression has a large impact on the in-plane phonons, not only shifting the frequencies significantly, but also changing the sign of the shift of the in-phase vibrations. On the other hand, the Bernal stacked graphite behaves more reasonably, that the frequencies of the in-phase modes (E$_{1u}$) increase while the out-of-phase (E${_2g}$) decrease with increasing interlayer coupling. We would like to point out again that no $sp^2$ to $sp^3$ transition occurs over the presented pressure range.

\section*{Discussions}
We now know that both the A-A and Bernal stacked graphene layers are very soft to compress, and under compression the electrons are not squeezed through the $sp^2$ network. The results that the stacking order has very small effect on the out-of-plane stiffness and interlayer interaction suggest that electrons distribution becomes uniform between layers under compression at various stacking orders, likely due to the electrons filling the area near the carbon hexagon centre of low electronic density. We would reasonably think that it is the $\pi$-electrons do the filling in the Bernal stacked graphene layers. On the other hand, the dramatic impact of out-of-compression on the in-plane phonon frequencies in the A-A stacked graphene layers strongly indicates that the $sp^2$-electrons are also involved. The $sp^2$-electrons filling the low-density area will cause a decrease of the overlap of the electronic orbitals of neighbouring carbon atoms and therefore result in a decrease of the in-plane phonon frequency as the calculations show. When we compress the graphene layers further, after the low-density area is filled, the in-plane phonon frequency will then increase, again just as the calculations show (change of the sign of the in-plane phonon shifts). We find an early published work indirectly supports this interpretation. It reported that the $sp^2$ to $sp^3$ transition of graphite is insensitive to the stacking order.\cite{Fahy87} It is insensitive because the out-of-plane stiffness and the interlayer interaction in different stacking orders are similar, as it becomes mainly uniform electronic distribution between graphene layers under compression.

\begin{figure}
	\includegraphics[width=\columnwidth]{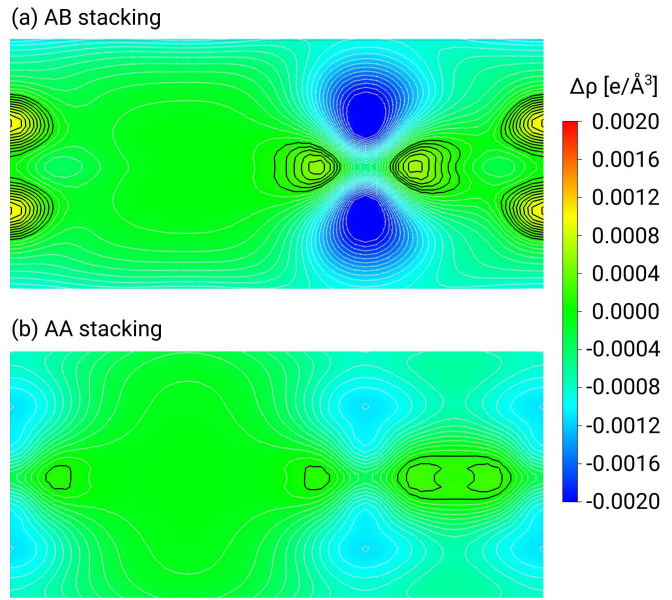}
	\caption{The charge density of the unstrain bi-layer graphene is subtracted by that of the compressed, and it is then plotted for (a) the Bernal and (b) A-A stacking. The colour scale is labelled. The contours are separated by 0.0001 e/\AA. The (110) graphene plane through the carbon atoms is plotted in the middle of each plot. We plot the charge density about 1 {\AA} above and below the graphene plane.}
	\label{cd}
\end{figure}

To illustrate this interpretation, we plot the charge density (see the supporting information), which determines all the presented results in this work, of the bi-layer graphene of the A-A and Bernal stacking. We focus on the graphene plane where the disruption of the C-C bonding is. Comparing the charge density of the unstrained and compressed bi-layer, we find that the difference in the charge density that causes such a large disruption in the $sp^2$ network as shown in Fig. \ref{biph}, is too tiny to be directly seen. Here we overlap the graphene plane and plot the difference in charge density between the unstrained and compressed bi-layer graphene, of Bernal and A-A stacking in Fig. \ref{cd}. The blue colour shows the increase of charge density under compression. The overlap of the $\pi$-electronic orbitals are clearly seen while in the A-A stacking, the $sp^2$ electrons `escape' out of the plane as the yellowish colour turns greenish, in the middle of the nearest C-C along the \textit{c}-axis. 

\section*{Conclusions}
In conclusion, we employed DFT to investigate the behaviour of the $\pi$-electrons of the graphene layers under compression, by obtaining the out-of-plane stiffness, the interlayer interaction, and the in-plane phonon frequencies. We find that the electrons can be hardly squeezed through the $sp^2$ network. Despite being slightly different, the out-of-plane stiffness and the interlayer interaction, both of which are mainly determined by the $\pi$-electrons of graphene layers, are very similar in both A-A and Bernal stacking. On the other hand, the shift under out-of-plane compression of the in-plane phonons of the A-A stacked graphene layers is significantly different from the reasonable shift of the Bernal stacking. Both the small effects on the out-of-plane properties and the large effects on the in-plane properties of the stacking order are surprising. We propose an interpretation, that electrons fill the centre areas of the carbon hexagons of low electronic density under compression, and form quite uniform electrons distribution between the graphene layers. In particular, in the A-A stacked graphene layers, the $sp^2$ electrons also contribute to the filling, inducing a softening of the C-C bond when the compression starts. This work strongly suggest not to ignore 3D effects, such as of out-of-plane compression, on a 2D material.

\bibliography{apssamp1}

\end{document}